\documentclass[num-refs]{nbdt-article}

\usepackage{siunitx}

\papertype{Original Article}
\paperfield{Journal Section}

\title{Moving outside the lab: The viability of conducting sensorimotor learning studies online}

\abbrevs{ABC, a black cat; DEF, doesn't ever fret; GHI, goes home immediately.}

\author[1, 2]{Jonathan S. Tsay}
\author[1, 2]{Alan S. Lee}
\author[1, 2]{Richard B. Ivry}
\author[1, 2]{Guy Avraham}

\affil[1]{Department of Psychology, University of California, Berkeley, CA 94706}
\affil[2]{Helen Wills Neuroscience Institute, University of California, Berkeley}

\corraddress{Jonathan S. Tsay, Department of Psychology, University of California, Berkeley, CA 94706, USA}
\corremail{xiaotsay2015@berkeley.edu}

\presentadd[\authfn{2}]{Department of Psychology, University of California, Berkeley, CA 94706, USA}

\fundinginfo{NIH, NINDS, Grant/Award Number: NS116883, NS105839, and NS1058389; Gorilla Experiment Builder; Foundation for Physical Therapy Research, PODSII scholarship}

\runningauthor{Tsay et al.}

\begin{document}

\maketitle

\begin{abstract}
Collecting data online via crowdsourcing platforms has proven to be a very efficient way to recruit a large and diverse sample. Studies of motor learning, however, have been largely confined to the lab due to the need for special equipment to record movement kinematics and, as such, are typically only accessible to specific participants (e.g., college students). As a first foray to make motor learning studies accessible to a larger and more diverse audience, we developed an online, web-based platform (OnPoint) to collect kinematic data, serving as a template for researchers to create their own online sensorimotor control and learning experiments. As a proof-of-concept, we asked if fundamental motor learning phenomena discovered in the lab could be replicated online. In a series of three experiments, we observed a close correspondence between the results obtained online with those previously reported from research conducted in the laboratory. This web-based platform paired with online crowdsourcing can serve as a powerful new method for the study of motor control and learning.

% Please include a maximum of seven keywords
\end{abstract}

\section{Introduction}
The ability to produce a wide repertoire of movements, and to adjust those movements in response to changes in the body and environment, is a core feature of human competence. This ability helps a tired ping-pong player compensate for her fatigue, and facilitates a patient’s motor recovery from neurological injury \cite{Tsay2020-nn, Krakauer2019-nm, Roemmich2018-ts}. By improving our understanding of how movements are learned, we can uncover general principles about how the motor system functions and develops, optimize training techniques for sport and rehabilitation, and design better brain-machine interfaces.

A paradigmatic approach for studying motor learning is to introduce a new mapping between the motion of the arm and the corresponding visual feedback \cite{Krakauer2000-ap}. Historically, such visuomotor perturbations were accomplished by the use of prism glasses that distort the visual field \cite{Helmholtz1924-ks}. Nowadays, virtual reality setups allow more flexible control over the relationship between hand position and a feedback signal \cite{Ghilardi2000-sn, Krakauer2000-ap, Krakauer2005-fe}. 

A commonly used perturbation is visuomotor rotation. Here, participants reach to a visual target with vision of the arm occluded. Feedback is provided in the form of a cursor presented on a computer monitor. After a brief training period during which the feedback corresponds to the actual hand position, participants see the cursor rotated by a certain amount (e.g., 15\textdegree{} clockwise rotation) with respect to the actual hand position. To nullify the mismatch (or error) between the expected position of the feedback (e.g., at the target location) and the actual position of the visual feedback, participants move in the opposite direction of the rotation. If the error is small, this change in heading direction will occur gradually over trials and occur outside the participant’s awareness, a phenomenon known as implicit sensorimotor adaptation. If the error is large, this implicit learning process may also be accompanied by more explicit and abrupt volitional adjustments in aiming \cite{Kim2020-cm, McDougle2019-fd, Shadmehr2010-rc}. While both implicit and explicit learning processes serve to nullify the visuomotor error, they show a number of distinct features: Not only do they occur at different rates, but the degree of explicit aiming flexibly scales with the size of the error whereas the maximum extent of implicit adaptation is invariant \cite{Bond2015-iu, Kim2018-gn, Morehead2017-bf}. 

Motor learning experiments are typically run in-person, exploiting finely calibrated apparatuses (digitizing tablets, robotic manipulandum, full VR displays, etc.) that provide data with high temporal and spatial resolution \cite{Ghilardi2000-sn, Krakauer2000-ap}. However, these experiments come at a cost: Beyond the expenses associated with purchasing specialized equipment, recruiting participants and administering the experiment can be slow and laborious, especially since testing is usually limited to one participant at a time. In-person participants are also mostly WEIRD (white, educated, industrialized, rich, and democratic), which may limit how the results of our research generalize to the population writ large \cite{Henrich2010-sd}. Moreover, exceptional circumstances that limit in-person testing, such as a global pandemic, can halt research progress \cite{Fauci2020-mk}. 

Social and cognitive sciences, on the other hand, have embraced online studies as a powerful alternative approach for collecting data for behavioral experiments \cite{Anwyl-Irvine2020-sn, Johnson2021-hz}. Researchers can recruit a large number of participants and perform rapid pilot testing on crowdsourcing platforms, such as Amazon Mechanical Turk (mturk.com) and Prolific (www.prolific.co). Compared to in-person studies, online studies not only reach a more diverse and representative set of participants \cite{Paolacci2014-mc}, but can also be accessible to individuals who have mobility issues that limit their capacity and willingness to come into the lab. 

Many in-person studies in the social sciences have been replicated online (e.g., \cite{Crump2013-wb}). In the present project, we set out to ask if sensorimotor learning studies can also be performed online. The loss of experimental control with online data collection may be especially problematic for kinematic data: Not only will the response devices (e.g., mouse or trackpad) be variable, but the setting in which movements are produced (e.g., focused or distracted environment) would be difficult to control. While previous efforts examining motor learning “in the wild” have advanced our understanding of motor learning \cite{noauthor_undated-vc, Crocetta2018-pd, Fernandes2011-cp, Ferrea2021-ye, Haar2020-gu, Kahn2018-md, Krakauer2020-tj, Lynn2020-yk, Takiyama2016-sx}, these studies have tended to focus on testing specific hypotheses (but see \cite{Coltman2021.06.14.448375} - a recently published study), and their online findings not readily comparable to those obtained in the lab. 

To explore the viability of online sensorimotor learning studies, we created a general-purpose online platform (OnPoint), one designed to be readily adopted by other research labs. This platform can be easily implemented on a standard laptop computer \cite{Tsay2020-mr}. As a proof-of-concept study, we asked whether data obtained via the platform replicates core motor learning and kinematic phenomena reported in previous in-person studies. Specifically, we focused on visuomotor rotation studies that elicit both implicit and explicit learning processes, and how these processes vary with the size of the rotation. Experiment 1 used a task that taps into both implicit and explicit processes involved in error-based motor learning, whereas Experiments 2 and 3 involved tasks that isolate the implicit process. We opted not to impose tight constraints over equating experiment methodologies (e.g., display size, response device) and demographics (e.g., age). Instead, we embraced the potential messiness expected from online studies, asking whether motor learning behavior observed online could still replicate those obtained in-person in the absence of experimenter oversight and demographic homogeneity. 

\section{Results}
\subsection{Experiment 1: Learning visuomotor rotation of different sizes}

Motor learning is frequently treated as an implicit phenomenon. Indeed, expert performers frequently comment on letting their “body do the thinking” when they execute an over learned skill \cite{Schmidt1987-bo}. However, these experts are also able to make rapid and flexible motor corrections, suggesting that even when behavior seems automatic, there remains considerable cognitive control \cite{Fitts1979-sp}. 

Recent work has highlighted how performance in even simple sensorimotor adaptation tasks reflects the operation of multiple learning processes that may solve different computational problems \cite{Benson2011-nw, Diedrichsen2010-mf, Haith2015-rz, Hegele2010-vd, Leow2018-fl, Miyamoto2020-fm, Taylor2014-xk, Werner2015-ya}. One source of evidence for this comes from a study by Bond and Taylor \cite{Bond2015-iu} who studied how people respond when the visual feedback associated with a movement is rotated, and in particular, when the size of the rotation varied in angular size. Explicit strategy use, as measured by verbal aim reports, was dominant when the error size was large, producing deviations in hand angle that scaled with the size of the perturbation. Yet, implicit adaptation, as measured by the aftereffects produced during a no-feedback block introduced immediately after learning, remained constant following perturbations ranging from 15\textdegree{} to 90\textdegree{}. 

In our first online experiment we varied the size of a visuomotor rotation, choosing a range based on Bond and Taylor (2015). We expected that during acquisition, the behavioral change would scale with rotation size, an effect we assumed would reflect the joint contribution of explicit and implicit processes. In contrast, we expected the aftereffect, a measure that should solely reflect implicit adaptation, would remain constant across rotation sizes. While Bond and Taylor had participants report their aiming location prior to each reach, we opted not to obtain aiming reports in this first proof-of-concept study, thinking that aim reports would add a level of procedural complexity (i.e., every reach would be accompanied by a key press for aim reports). Nonetheless, previous work indicates that the absence of aim reports neither significantly impacted the scalar signature of explicit aiming \cite{Neville2018-uf, Taylor2014-xk} nor the invariant nature of implicit adaptation \cite{Kim2018-gn, Morehead2015-re}. 

After a series of baseline blocks to familiarize the participants with the apparatus and basic trial procedure, the participant experienced one of four rotation sizes (15\textdegree{}, 30\textdegree{}, 60\textdegree{}, 90\textdegree{}), with the perturbation constant for an entire block of 80 rotation trials (Figure 1a). Participants were instructed to hit the target with their cursor. The learning functions are presented in Figure 1b, alongside the in-person results from Bond and Taylor (Figure 1d) as a point of comparison. In both online and in-person settings, learning scaled with the size of the rotation during early learning (main effect of perturbation size: F(1,136)=64.5,p<0.01) and reached an asymptote close to the size of the perturbation (late learning, main effect of perturbation size: F(1,136)=810.1,p<0.01). Hand angle dropped dramatically in the no-feedback aftereffect block although it remained greater than zero, and in the direction away from the feedback (all groups: p<0.01), the signature of an implicit aftereffect. Critically, the magnitude of the aftereffect did not scale with the size of the rotation (main effect of perturbation size: F(1,136)=2.5,p=0.12). This overall pattern is consistent with the hypothesis that during learning, implicit adaptation was supplemented by strategic aiming to negate the error introduced by the rotation. During the aftereffect phase, participants ceased to aim, revealing the state of implicit adaptation, an effect that was independent of the size of the perturbation.

\begin{figure}[!htb]
\centering
\includegraphics[width=14cm]{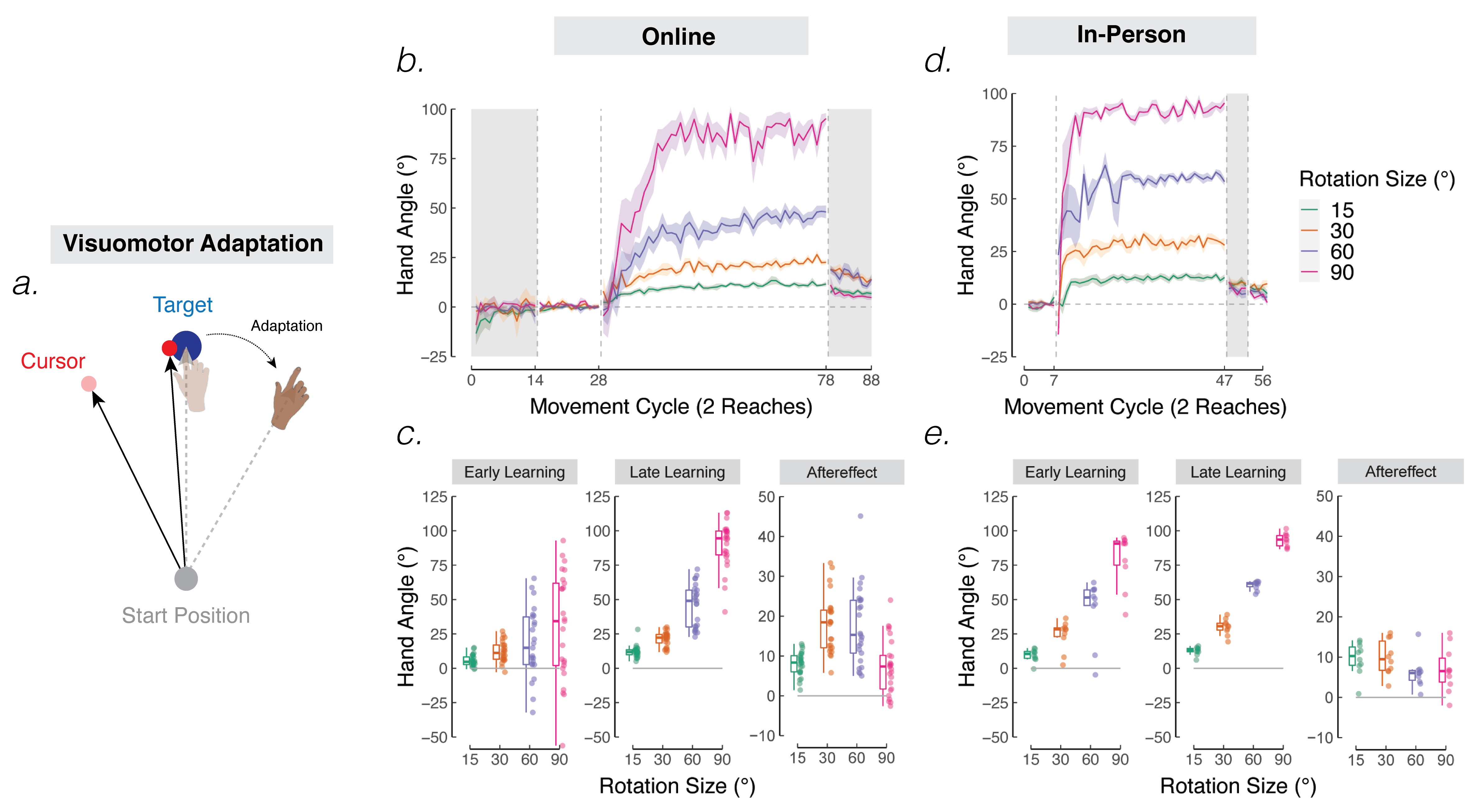}
\caption{\textbf{Sensorimotor learning in response to visuomotor rotations.} \textbf{(a)} Schematic of a visuomotor rotation task. The cursor feedback (red dot) was rotated with respect to the movement direction of the hand, with the size of the rotation varied across groups (15\textdegree{}, 30\textdegree{}, 60\textdegree{}, or 90\textdegree{}). Translucent and solid colors display hand and cursor positions at early and late stages of learning, respectively. \textbf{(b, d)} Mean time courses of hand angle for 15\textdegree{} (green), 30\textdegree{} (yellow), 60\textdegree{} (purple), and 90\textdegree{} (pink) rotation conditions from the online and in-person experiment of Bond and Taylor (2015). Hand angle is presented relative to the target (0\textdegree{}) during veridical feedback, no-feedback (grey background), and rotation trials. Shaded region denotes SEM. Note that Bond and Taylor (2015) used eight target locations, and thus, had 8 reaches per cycle. However, we limited our analysis of Bond and Taylor (2015) to the same two targets used in the online variant (i.e., 45\textdegree{} and 135\textdegree{} targets). \textbf{(c, e)} Average hand angles during early and late phases of the rotation block, and during the no-feedback aftereffect block from the online \textbf{(c)} and in-person \textbf{(e)} experiments. Box plots denote the median (thick horizontal lines), quartiles (1st and 3rd, the edges of the boxes), and extrema (min and max, vertical thin lines). The data from each participant is shown as translucent dots.}
\end{figure}

As a secondary question, we asked whether online and in-person data were nominally similar. We recognize the limitations of directly comparing absolute values between settings given numerous methodological differences (e.g., setting and presence/absence of aiming reports). Nonetheless, the degree of overall adaptation was similar between settings for all phases of learning (main effect of setting: Early F(1,136)=0.5,p=0.46; Late F(1,136)=1.3,p=0.24; Aftereffect: F(1,136)=1.9,p=0.17). 

However, there were a few differences: First, the adaptation rate for the online participants was slower early in learning (interaction of rotation size x setting: F(1,136)=64.5,p<0.01), an effect neither present during late learning (F(1,136)=0.37,p=0.54) nor during the aftereffect phase (F(1,136)=0.1,p=0.75). The early learning difference likely reflects the inclusion of the aiming report component in Bond and Taylor (2015). The requirement to report the aiming location as well as the presence of visual landmarks provided for these reports likely helped participants identify an appropriate aiming strategy \cite{Taylor2014-xk, Maresch2020-nv}. Given that the online sample was older than that of in-person, age may also have been a contributing factor \cite{Hegele2010-vd, Vandevoorde2019-af, Vandevoorde2020-dz, Wolpe2020-un}. Second, the online data were more variable, either measured in terms of within-participant variability of heading angle (online SD: 19.0 $\pm$ 1.5\textdegree{} vs in-person SD: 12.4 $\pm$ 1.3\textdegree{}; p<0.01) or between-participant variability (see Table 1 in \emph{Supplementary Tables}). Greater variability in the online experiment may be attributed to differences in the type of movement (wrist vs arm), sources of distraction (presumably greater for online), and/or demographic differences. Third, the mean hand angle in the aftereffect phase for online participants (Figure 1b \& 1c) exhibited a small, non-monotonic variation. Although we do not have a sensible working hypothesis for this effect, we note that the aftereffect data do not show monotonic scaling with rotation size, similar to that observed in in-person studies.

\subsection{Experiment 2: Adaptation in response to non-contingent rotated visual feedback}

In Experiment 2, we turn to a method designed to measure implicit learning in the absence of strategic aiming. Motivated by the idea that adaptation is obligatory in response to a visual sensory prediction error \cite{Morehead2017-bf}, Morehead et al (2017) replaced the standard movement-contingent feedback cursor with a “visual clamp”. Here, the cursor follows an invariant trajectory on all trials, with the radial position dependent on the participant’s hand position (as in standard feedback), but the angular position always shifted from the target by a fixed angle (Figure 2a). In this manner, the angular position of the cursor is no longer contingent on the participant’s movement. Despite instructions that explain the manipulation and emphasize that the participant should ignore the cursor and always move directly to the target, this manipulation induces gradual changes in hand angle away from the target in the direction opposite to the perturbation. Learning here is entirely implicit, verified in both subjective interviews provided by participants at the end of the experimental session \cite{Avraham2021-cv, Kim2018-gn, Kim2019-xa}, as well as in reports of sensed hand location obtained on probe trials throughout the adaptation block \cite{Tsay2020-me}. 

The clamp method provides another way to ask how error size influences implicit adaptation. Morehead et al (2017) demonstrated that the rate of adaptation is largely invariant over a wide range of error sizes (clamp angles ranging from 7.5\textdegree{} - 95\textdegree{}). Moreover, the asymptote has also been shown to be independent of the error size for this range of perturbations, averaging between 15\textdegree{} - 25\textdegree{} across several studies \cite{Avraham2020-zd, Kim2018-gn, Tsay2020-km, Kim2019-xa}. 

Experiment 2 used a design based on a subset of the conditions in Morehead et al (2017). We examined adaptation in response to visual clamps of 7.5\textdegree{}, 15\textdegree{}, and 30\textdegree{} with each perturbation tested in separate groups of participants as in Experiment 1. We also included a 0\textdegree{} condition, one in which the cursor feedback always moved directly to the target. This condition provides a baseline to ensure that changes in hand angle in the other groups are driven by error-based learning, rather than changes due to fatigue or proprioceptive drift \cite{Brown2003-br, Brown2003-qc, Cameron2015-sw, Wann1992-ai}. 

These learning functions are presented in Figures 2b (online) \& 2d (in-person). We analyzed our data together with those obtained by Morehead et al (2017), evaluating mean performance at three phases of the experiment: Early adaptation, late adaptation, and aftereffect. As expected, there was no consistent change in performance in response to the 0\textdegree{} clamp in our data (one sample permutation test: early learning: p=0.62; late adaptation, p=0.87; aftereffects, p=0.19), similar to that observed in Morehead et al (2017) (one sample permutation test: early learning, p=0.46; late adaptation, p=0.26; aftereffects, p=0.82). In contrast, adaptation was evident in all stages of learning for the non-zero clamps (one sample permutation test, all p<0.05). 

\begin{figure}[bt]
\centering
\includegraphics[width=14cm]{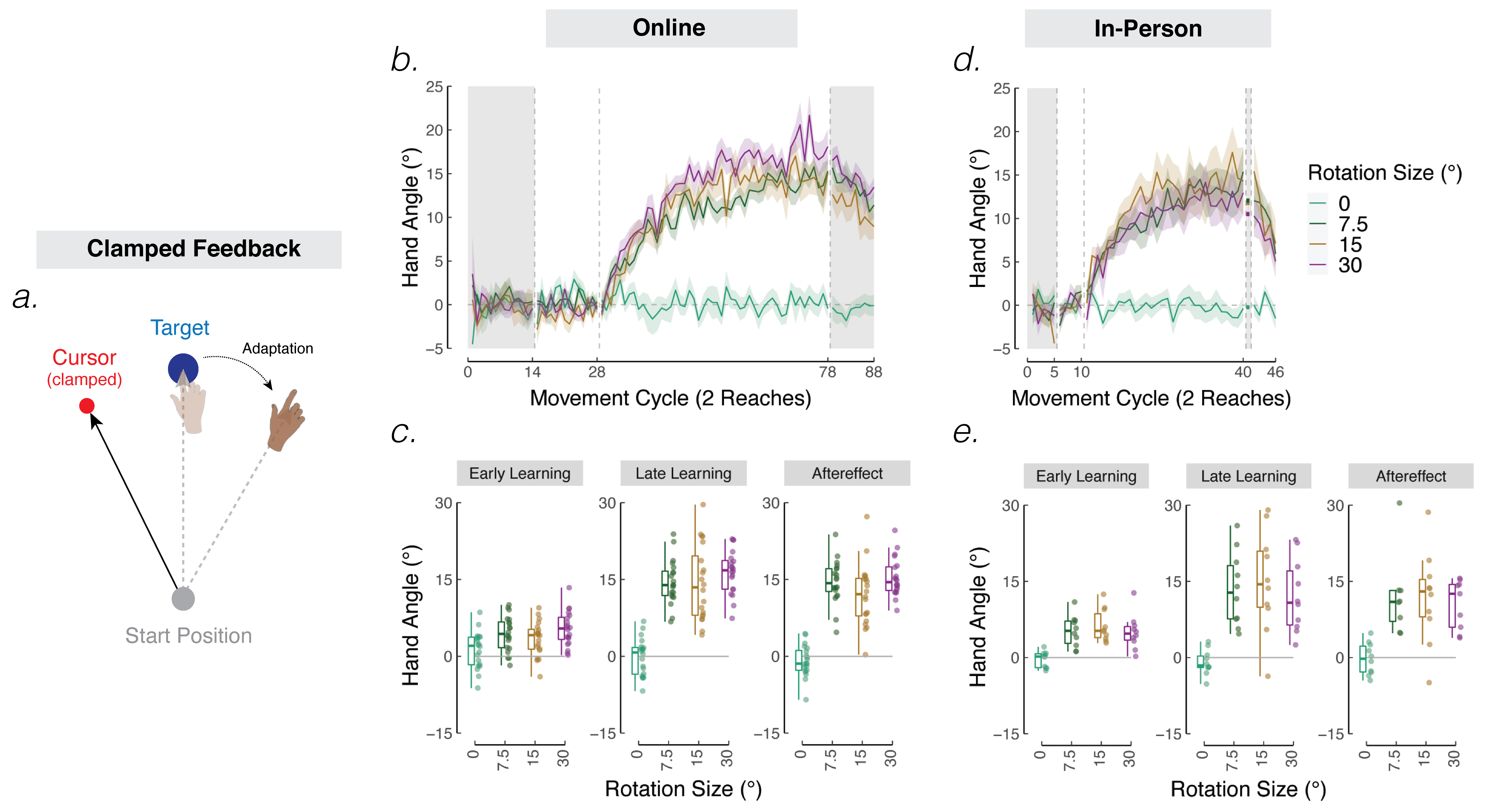}
\caption{\textbf{Sensorimotor adaptation in response to non-contingent displaced visual feedback.} \textbf{(a)} Schematic of the clamped feedback task. The cursor feedback (red dot) follows a trajectory rotated relative to the target, independent of the position of the participant’s hand. The rotation size remains invariant throughout the rotation block but varied across groups. Participants were instructed to move directly to the target (blue circle) and ignore the visual feedback. The translucent and solid colors display hand position early and late in learning, respectively. \textbf{(b, d)} Mean time courses of hand angle for 0\textdegree{} (green), 7.5\textdegree{} (dark green), 15\textdegree{} (brown), and 30\textdegree{} (dark purple) rotation conditions (in-person experiment, adapted from Morehead et al. 2017). Hand angle is presented relative to the target (0\textdegree{}) during no-feedback (dark grey background), veridical feedback, and rotation trials. Shaded region denotes SEM. Note that Morehead et al (2017) used eight targets. However, we limited our analysis of Morehead et al (2017) to the same two targets used in the online variant (i.e., 45\textdegree{} and 135\textdegree{} targets). \textbf{(c, e)} Average heading angles during early and late phases of the rotation block, and during the no-feedback aftereffect block from the online \textbf{(c)} and in-person \textbf{(e)} experiments. Box plots denote the median (thick horizontal lines), quartiles (1st and 3rd, the edges of the boxes), and extrema (min and max, vertical thin lines). The data from each participant is shown as translucent dots.}
\end{figure}

For non-zero clamp sizes, adaptation in both experiments did not scale with rotation size during early learning (main effect of rotation size: F(1,86)=0.2,p=0.66), late learning (F(1,86)=0.0,p=0.91), and the no-feedback aftereffect block (F(1,86)=0.0,p=0.96) (Figures 2c and 2e). The functions for the 7.5\textdegree{}, 15\textdegree{}, and 30\textdegree{} clamps reach an asymptote of about 15\textdegree{}, with the range of values across individuals similar to that seen in the aftereffect data of Experiment 1. We note that the magnitude of adaptation is approximately twice that of the perturbation for the 7.5\textdegree{}. While this might seem puzzling, it is important to keep in mind that, unlike normal adaptation studies where the position of the feedback cursor is contingent on the hand movement and thus, the size of the visual error is reduced throughout adaptation, the error size remains invariant with the clamped feedback task and continues to drive adaptation. 

In terms of a comparison to in-person results, the online data were similar to those collected by Morehead et al. (no main effect of setting: Early, F(1,86)=0.8,p=0.13; Late,  F(1,86)=0.2,p=0.63, Aftereffects, F(1,86)=0.0,p=0.98). Within-participant variability was again greater in the online group (In-person SD: 4.5 $\pm$ 1.4\textdegree{}; Online SD: 8.2 $\pm$ 0.4\textdegree{}; p<0.01), but between-participant variability was surprisingly similar between online and in-person groups (see Table 2 in \emph{Supplementary Tables}). 

In sum, the online results in Experiment 2 replicate two core insights derived from in-person studies using clamped feedback: First, adaptation occurs implicitly and automatically in response to a visual sensory prediction error. Second, the learning function is invariant across a large range of error sizes, both in the shape of the function and its asymptotic value. 

\subsection{Experiment 3: Adaptation in response to variable, non-contingent rotated visual feedback}

The use of a fixed perturbation for each participant in Experiments 1 and 2 allowed us to assess the full learning curve and aftereffect. This design often lacks the power to identify subtle differences in sensitivity to error size because the standard methods of analysis involve smoothing the data over multiple trials and making comparisons between individuals (or across sessions if a repeated measures design is employed). An alternative approach to study the effect of error size on implicit adaptation is to use a random perturbation schedule, exposing each individual to a range of error sizes throughout the perturbation block. By including both clockwise and counterclockwise rotations, there is no cumulative measure of learning; rather, the analysis focuses on trial-to-trial changes in heading angle (Figure 3a) \cite{Avraham2020-zd, Hutter2018-kh, Kording2004-ru, Marko2012-oz, Tsay2020-km, Wei2009-ix, Wei2010-gi}. 
 
Following the in-person method used in Tsay et al (2021) \cite{Tsay2021.06.20.449180}, we varied the size of the non-contingent clamped feedback across trials. Each participant was exposed to a set of eight rotation sizes between 0\textdegree{} - 60\textdegree{} with four of these involving clockwise rotations and the other four involving counterclockwise rotations of the same size. To sample a large range while keeping the experiment within 1 hour, participants received different sets of perturbations (total of four sets, see \emph{Methods}). Given that the eight perturbations within a set have a mean of zero, learning should not accumulate across trials. Similar to Experiment 2, participants were instructed to ignore the cursor feedback and always move directly to the target. 

As a trial-by-trial measure of implicit adaptation, we averaged each participant’s change in hand angle from trial n to trial n + 1, as a function of the rotation size on trial n. As can be seen in Figure 3b, the participants showed a sign-dependent change in hand angle in response to the clamped feedback, similar to that observed in the in-person study of Tsay et al (2021) (Figure 3c). The function is sublinear, composed of a quasi-linear zone for smaller perturbations (up to around 16\textdegree{}) and a saturation range for larger perturbations. In both the online and in-person studies, the mean changes in hand angle fall within a similar range ($\pm$2.5\textdegree{}). 

To statistically evaluate these data, we first calculated the slope from each individual’s learning function. The slopes were significantly less than 0 for the online and in-person experiments (both p<0.01), confirming robust sign-dependent implicit adaptation. We then asked whether the learning functions were sublinear by comparing, for each individual, the slope when computed using all perturbation sizes to the slope when using only the small perturbations (in-person: $\pm$0\textdegree{}, $\pm$4\textdegree{}; online: the smallest two rotation sizes in their set, Set 1: $\pm$2\textdegree{}, $\pm$4; Set 2: $\pm$10\textdegree{}, $\pm$25\textdegree{}; Set 3: $\pm$7.5\textdegree{}, $\pm$15\textdegree{}; Set 4: $\pm$2\textdegree{}, $\pm$4\textdegree{}). If the function is sublinear, the absolute slope calculated using all of the rotation sizes should be smaller (less negative). The results indicated that the functions were sublinear in both sets of data (in-person, p=0.01; online, p=0.02). 

We also compared movement variability between settings. Specifically for a given clamp size, we asked whether the within-subject standard deviation of the $\Delta$ hand angle differed between settings. We limited our analysis to the most comparable clamp sizes between settings: 4\textdegree{} online vs 4\textdegree{} in person, and 15\textdegree{} online (closest comparable rotation size) vs 16\textdegree{} in person. As observed in Experiments 1 and 2, within-subject variability was larger in the online group for both clamp sizes (4\textdegree{} vs 4\textdegree{}: p<0.001; 15\textdegree{} vs 16\textdegree{}: p<0.001) compared to the in-person group. Similarly, for these same clamp sizes, there was also larger between-participant variability in the online group compared to the in-person group (see Table 3 in \emph{Supplementary Tables}). 

In sum, the results of Experiment 3 show a striking correspondence to that obtained in-person using a near-identical design. Moreover, the functions, both in shape and magnitude are quite similar to that reported in previous studies that have used a variable-sized perturbation to study implicit adaptation \cite{Kasuga2013-bq, Ranjan2020-zs,Tsay2020-km, Wei2009-ix, Wei2010-gi}.

\begin{figure}[!htb]
\centering
\includegraphics[width=14cm]{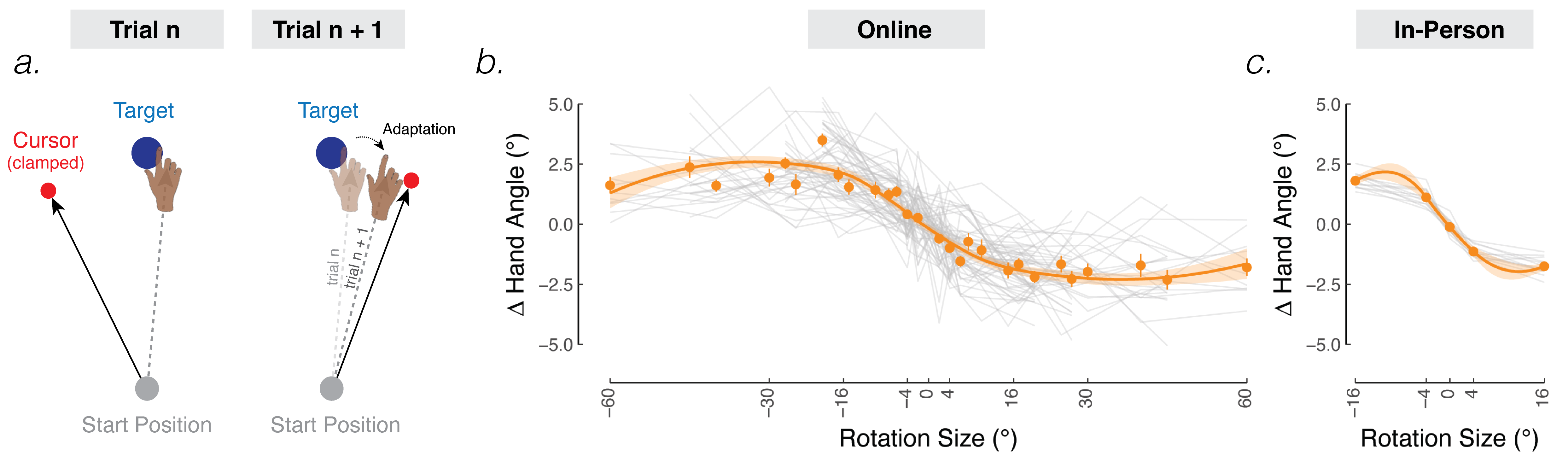}
\caption{\textbf{Trial-by-trial estimates of sensorimotor adaptation in response to variable, non-contingent visual feedback.} \textbf{(a)} Schematic of the task. The cursor feedback (red dot) was rotated relative to the target, independent of the position of the participant’s hand. The size of the rotation was varied randomly on a trial-by-trial basis. \textbf{(b, c)} The average change in hand angle from trial n to trial n + 1 is plotted as a function of rotation size on trial n. Thin grey lines are individual data collected online \textbf{(b)} and in-person \textbf{(c)} (figure adapted from Tsay et al (2021), see Experiment 3 in-person procedure in the \emph{Methods}), with the best-fitting loess line indicated by the orange curve (shaded region denotes SEM). Orange points denote group means and bars denote SEM.}
\end{figure}

\section{Discussion}

An online platform for conducting motor learning experiments offers a new research tool to collect kinematic data from large and diverse samples in an efficient and low-cost manner. As a proof-of-concept, we reported three experiments that examined behavioral changes in response to perturbed visual feedback, adopting established tasks for our online platform. Qualitatively, the results from all three online studies show a close correspondence with those obtained from in-person studies. Specifically, early and late learning scaled with the size of the rotation when both implicit and explicit processes were involved (Exp 1), but implicit adaptation was insensitive to error size across a range of large errors (7.5\textdegree{} - 90\textdegree{}, Exps 1 and 2). In a more granular analysis, we found that implicit adaptation in both in person and online studies was selectively sensitive to smaller errors (Exp 3), showing a saturated response to large errors. Together, these results demonstrate that online experiments provide a viable alternative to study sensorimotor adaptation outside the confines of the traditional laboratory setting. 

\subsection{Shared and distinct features of online and in-person experiments}

The similar motor adaptation behavior observed between settings is especially striking in light of significant differences between the two platforms, three of which we highlight here. First, in-person participants typically use their entire arm to control a robotic manipulandum or a digitizing pen on a large tablet surface. For the online studies, the participants mostly used a trackpad (something we encouraged but have also found to the norm in unrestricted online studies). Although we did not obtain detailed reports, we assume that “reaching” movements with the trackpad involved relatively small rotations about the wrist, perhaps coupled with extension of the index finger. These types of movements will entail a different set of biomechanical and sensory constraints compared to reaches performed by moving the entire arm \cite{De_Rugy2009-lz, Debats2018-bx, Hollerbach1982-ps, Yin2019-vy}. Nonetheless, adaptation appears to be quite similar for these different classes of movements \cite{Fernandes2011-cp, Takiyama2016-sx}. Indeed, in an unpublished study using a standard 45\textdegree{} visuomotor rotation, we did not find qualitative behavioral differences in measures of adaptation for mouse users (i.e., arm movements) vs trackpad users (i.e., finger movements) (Fig 4, see \emph{Supplementary Figure}). 

Second, while in-person experiments typically involve constant experimenter supervision and verbal instruction to ensure the movements are performed properly, online experiments involve no experimenter supervision and only a limited set of instructions (e.g., feedback messages like “too far” or “too slow”). We did include several instructional videos to demonstrate important components of the task and a post-experiment survey to assess whether participants understood the task, but there remains limited control during online experimental sessions. To our surprise, minimal experimental oversight in the online setting did not swamp the data with noise. 

Third, whereas the hand is typically hidden from view and moving in-plane with visual feedback for in-person experiments, the hand is always visible and moving out of plane for online experiments. Visual feedback of the hand has been shown to modulate motor planning \cite{Sober2005-aq, Wilkie2012-pl} and can attenuate implicit adaptation \cite{Wong2019-yv}. However, implicit adaptation remains qualitatively similar between experimental settings. Perhaps online participants seldom look at their hand when they move since they are likely adept trackpad users who are very familiar with how movements of the finger on the trackpad map on to movements of the cursor on their vertical screen. 

\subsection{Moving forward with online experiments}

While the main features of performance and learning in our online experiments were quite similar to that reported in the comparison in-person experiments, there was one notable difference: Within- and between-participant variability was greater in the online group compared to in-person. As noted previously this increase in variability may be due to differences in the movements produced for online and in-person experiments, experimental supervision, participant demographics, or some combination of these and other variables. Given that these differences were observed across all phases of the experimental tasks, they are unlikely to limit our inferences regarding learning. Nonetheless, the consistent variability differences underscore the importance of ensuring that the sample size is tailored appropriately for online studies. 

Improvements in software and hardware may further close the gap between in-person and online studies. Specifically, online experiments present stimuli with greater and more variable temporal delays \cite{Anwyl-Irvine2020-jh}. This may be a critical limitation for some studies of adaptation given that the adaptation decreases when the feedback is delayed \cite{Brudner2016-uj, Held1966-vx, Kitazawa1995-rf, Dawidowicz2021.05.27.445962}. For this reason, we would urge caution in the use of online studies if the focus of the research is on absolute values. Concerns with temporal delays are mitigated for relative comparisons (such as the analyses presented here to compare conditions in the online studies). 

Alternatively, instead of fully replacing in-person studies with those administered online, the future might be one in which researchers employ a hybrid approach, running online experiments when there is no need for specialized lab equipment and conducting in-person experiments when there is (e.g., robot to introduce force-field perturbation or device to vibrate tendons to disrupt proprioception) \cite{Johnson2021-hz}. This hybrid approach has shown promise in our own work involving populations with mobility deficits \cite{Saban2021-am}. Special equipment could also be sent to participants’ homes, accompanied by detailed set-up instructions. Such equipment can include virtual reality headsets \cite{Brookes2020-wf} and mobile systems for neural recording \cite{Askamp2014-ya, Gilron2021-hn}.

\subsection{Moving forward with the OnPoint platform}

The OnPoint platform is a living code, one that will be continuously updated by ourselves and members of the motor control community. In the experiments reported here, we did not record the participant’s screen size. However, we have recently implemented a new version of OnPoint that records screen size and thus allows us to determine the exact size of the stimuli. 

In addition, we did not standardize the sensitivity (e.g., gain) of the participant’s response device. This parameter is notoriously difficult to control, given that different devices have different sensitivity default settings and users tailor these settings to their idiosyncratic preferences. Device sensitivity will affect the movement distance and this could influence adaptation \cite{Wei2009-ix}. We expect this variability did not distort the current results given the relatively large sample size per group and fairly small variation in trackpad size across standard computers. However, there are likely significant differences in movement distance between online and in-person studies; as such, this issue must be kept in mind when making quantitative comparisons between the two experimental settings. We now include a step-by-step workflow in the GitHub template describing how to obtain device sensitivity. By following this procedure, researchers will be in the position to include device sensitivity as a covariate and even have the option to modify the instructions so that all participants use the same device sensitivity. 

In our post-experiment survey, participants reported difficulty returning to the center location even though we provided veridical feedback. We have recently implemented a new version of OnPoint that bypasses the need for a large return movement: Following the reach feedback, the visual cursor disappears and then re-appears at a random location near the start position (2 cm). Participants only need to make a small movement to get to the start location. This does mean that the joint angle at the start location will vary across trials. However, these small postural changes are unlikely to impact adaptation. Moreover, the ease with which participants find the start location with this new procedure (eliminating what has been the most frustrating part of the task) outweighs any putative influence of the return movement on adaptation \cite{Taylor2013-jr}.

Throughout our experience with the online experiments, we have gathered a set of suggestions that may improve the data quality for researchers interested in “moving outside the lab”. First, make your task fun and engaging. Tasks that capture the participant’s attention will likely result in higher quality data. Second, pilot, pilot, pilot! Send your experiment to colleagues to ensure that your instructions are clear and concise and get their valuable feedback. Third, include periodic attention and instruction checks so that you can identify distracted or less motivated participants. Whether participants are attentive can also be inferred through fluctuations/spikes in hand angle and other kinematic variables of interest. Fourth, pay your participants fairly. Although we have not done a systematic assessment, we strongly suspect that data quality improves if the participants are given reasonable financial incentives. 

In summary, online experiments provide a viable and novel way to test predictions about motor learning with large numbers of participants in a short amount of time. Whereas it would have taken months to collect the data reported here if the studies were run in-person, our online platform allowed us to collect these data within a few days. Moreover, participants recruited online represent greater diversity, one that spans a range in terms of age, ethnicity, handedness, and years of education (see \emph{Participants}) \cite{Paolacci2014-mc}. Most strikingly, despite numerous differences between in-person and online settings, many core phenomena central to our understanding of sensorimotor learning can be replicated online. This proof-of-concept study thus paves the way for taking sensorimotor research beyond the lab.

\section{Methods}

\subsection{Participants}
The protocol was approved by the institutional review board at the University of California, Berkeley. Participants (n = 260; age range = 19 – 65, mean age $\pm$ sd = 35.0 $\pm$ 10.2) were recruited from the Amazon Mechanical Turk (AMT) over 8 days. Participants received financial compensation for their participation at an \$8 per hour rate. Recruitment was restricted to the United States. 

Based on the participants who completed an optional online survey (n = 180 out of 260 responded, 130 declined to participate in the survey), there were 100 male participants, 69 female participants, and 11 identified as other. 124 of the participants identified as White, 17 as Asian, 25 as African American, 1 as a Pacific-Islander, 2 as multi-racial, and 11 declined to answer. 144 of the participants were right-handed, 22 left-handed, and 4 self-identified as ambidextrous. In terms of response devices, we encouraged participants to use a trackpad to limit device-related variance \cite{Moher2019-pj}, but did not enforce this requirement. As a result, there were 154 trackpad users but only 16 mouse users (others opted not to provide this information). No statistical methods were used to determine the target sample sizes. 

Limiting the time of data collection to regular business hours: Previous studies have shown that the demographic make-up of online participants varies across the time of day \cite{Arechar2017-lr, Casey2016-bt}. For instance, people who work in typical white-collar jobs would be unavailable to complete studies during regular business hours. As such, participants during regular working hours are primarily those who use AMT as a primary source of income. These “professional” participants typically make less errors and complete studies more efficiently \cite{Casey2016-bt, Paolacci2014-mc, Hauser2019-wo}. Thus, we limited the time of data collection to regular business hours by launching the experiment in the morning around 9 am California time. 

Rationale to compare online results with existing in-person studies, instead of running new within-participant studies: Due to a global pandemic \cite{Fauci2020-mk}, in-person research was halted. We therefore opted to compare our online results with existing published results in a post-hoc, between-participant manner, instead of performing entirely new experiments using a within-participant design (e.g., session 1 in-person $\rightarrow$ session 2 online). This is a proof-of-concept study for the viability of online motor learning experiment. Thus, we aimed to match the general protocol design of the online experiments to the in-person experiments that obtained the original results. Second, within-participant designs may result in possible transfer/order effects, which can be difficult to nullify even with counterbalancing. For instance, adaptation that incorporates explicit strategies is faster in a second session (also known as savings) \cite{Avraham2021-cv, Haith2015-rz, Herzfeld2014-bp, Leow2020-ve, Morehead2015-re}, and implicit adaptation has been shown to attenuate upon re-learning \cite{Avraham2021-cv, Stark-Inbar2017-zs, Tsay2021-qn, Yin2020-jj}. 

\subsection{Apparatus}
Participants used their own computer to access a dynamic webpage (HTML, JavaScript, and CSS) hosted on Google Firebase. The task progression was controlled by JavaScript code running locally in the participant’s web browser. We assumed that monitor sampling rates were typically around 60 Hz, with little variation across computers \cite{Anwyl-Irvine2020-jh}. The size and position of stimuli were scaled based on each participant’s screen size, which was automatically detected. 

\subsection{Reaching task Procedure} The participant made reaching movements by moving the computer cursor with the trackpad or computer mouse. The stimuli parameters reported below are for a typical monitor size of 13’’ (1366 x 768 pixels) \cite{Anwyl-Irvine2020-jh}. On each trial, the participants made a center-out planar movement from the center of the workspace to a visual target. The center position was indicated by a white circle (0.5 cm in diameter) and the target location was indicated by a blue circle (also 0.5 cm in diameter). The radial distance of the target from the start location was 6 cm. In experiments 1 and 2, the target could appear at one of two locations on an invisible virtual circle (45\textdegree{}: upper right quadrant; 135\textdegree{}: upper left quadrant). For these experiments, a movement cycle is defined as 2 consecutive reaches, one to each target. In Experiment 3, the target appeared in a single position at 45\textdegree{} throughout the entire experiment. 

To initiate each trial, the participant moved the cursor, represented by a white dot on their screen (0.5 cm in diameter), into the start location. Note that unlike in-person experiments where movements of the hand and cursor are in-plane (mostly in the horizontal plane), movement of the hand and cursor are out-of-plane (hand = horizontal plane; cursor = vertical plane). During this initialization phase, feedback was provided when the cursor was within 4 cm of the start circle. Once the participant maintained the cursor in the start position for 500 ms, the target appeared. The location of the target in Experiments 1 and 2 was selected in a pseudo-randomized manner. The participant was instructed to reach, attempting to rapidly “slice” through the target. The feedback cursor, when presented (see below) remained visible throughout the duration of the movement and remained fixed for 50 ms at the radial distance of the target when the movement amplitude reached 6 cm. If the movement was not completed within 500 ms, the message “too slow” was displayed in red 20 pt. Times New Roman font at the center of the screen for 750 ms. 

The feedback could take one of the following forms: veridical feedback, no-feedback, rotated contingent feedback (Exp. 1), and rotated non-contingent (“clamped”) feedback (Exps. 2 and 3). During veridical feedback trials, the movement direction of the visual feedback was veridical with respect to the movement direction of the hand. During no-feedback trials, the feedback cursor was extinguished as soon as the hand left the start circle and remained off for the entire reach. The (veridical) cursor only became visible during the return phase of the trial, when the cursor was within 4 cm of the start circle. We note that Bond and Taylor (2015) used a different method to re-center the cursor between trials: a visual ring that indicates the hands radial distance from the center. We opted not to use this method because, in a pilot experiment, the visual ring led to long search times and (we assume from our own experience) frustration with the task (something that can also occur in online experiments using the ring method). Use of veridical feedback does not seem to influence adaptation in the same hemifield (e.g., upper half of the screen) and thus should not influence behavior for all experiments presented here \cite{Taylor2013-jr}. 

With rotated contingent feedback, the cursor moved at an angular offset relative to the position of the hand; the radial position of the cursor corresponded to that of the hand up to 6 cm, at which point, the cursor position was frozen for 500 ms, after which the cursor disappeared. During rotated clamped-feedback trials, the cursor moved at a specified angular offset relative to the position of the target, regardless of the movement direction of the hand (“clamped feedback”); as with rotated contingent feedback, the radial position of the cursor corresponded to that of the hand.

\subsection{Experiment 1: Learning visuomotor rotation of different sizes} AMT participants (N = 100, mean age $\pm$ sd = 34.6 $\pm$ 9.0) completed a visuomotor adaptation task consisting of four blocks of trials (178 trials total: 89 trials x 2 targets): Baseline no-feedback block (28 trials), baseline feedback block (28 trials), rotated feedback block (100 trials), and no-feedback aftereffect block (20 trials). During the rotation block, each participant was assigned one of four rotation sizes (15\textdegree{}, 30\textdegree{}, 60\textdegree{}, 90\textdegree{}; 25 participants/group), with the direction of the rotation (clockwise or counterclockwise) counterbalanced across participants.

Prior to each baseline block, the instruction “Move directly to the target as fast and accurately as you can” appeared on the screen. Prior to the rotation block, a new instruction message was presented: “Your cursor will now be rotated by a certain amount. In order to continue hitting the target, you will have to aim away from the target.” Prior to the no-feedback aftereffect block, the participants were instructed “Move directly to the target as fast and accurately as you can.” 

\subsection{Experiment 2: Adaptation in response to non-contingent rotated visual feedback} A new sample of AMT participants (N = 80, mean age $\pm$ sd = 34.8 $\pm$ 10.2) completed a visuomotor adaptation task, with the same block structure as in Experiment 1 (178 total trials) with one critical difference: Rotated, non-contingent feedback was used during the rotation block, with the clamp fixed at one of four angular offsets relative to the target (0\textdegree{}, 7.5\textdegree{}, 15\textdegree{}, 30\textdegree{}; 20 participants/group). The direction of the non-zero clamps (clockwise or counterclockwise) was counterbalanced across participants. 

The instructions for baseline and no-feedback aftereffect blocks were identical to those used in Experiment 1. Prior to the rotation block, the instructions were modified to read: “The white cursor will no longer be under your control. Please ignore the white cursor and continue aiming directly towards the target.” To clarify the invariant nature of the clamped feedback, three demonstration trials were provided. On all three trials, the target appeared straight ahead (90\textdegree{} position) and the participant was told to reach to the left (demo 1), to the right (demo 2), and backward (demo 3). On all three of these demonstration trials, the cursor moved in a straight line, 90\textdegree{} offset from the target. In this way, the participant could see that the spatial trajectory of the cursor was unrelated to their own reach direction.

\subsection{Experiment 3: Adaptation in response to variable, non-contingent rotated visual feedback} A new sample of AMT participants (N = 60, age was not obtained due to an experimental error) completed a visuomotor adaptation task consisting of four blocks of trials (255 total trials): Baseline no-feedback block (5 trials), baseline feedback block (15 trials), rotated feedback block (230 trials), and no-feedback aftereffect block (5 trials). During the rotation block, the non-contingent feedback varied from trial to trial, both in direction (clockwise or counterclockwise) and angular offset. Participants were assigned one of four sets of rotation sizes (Set 1: $\pm$2\textdegree{}, $\pm$4\textdegree{}, $\pm$6\textdegree{}, $\pm$20\textdegree{}; Set 2: $\pm$10\textdegree{}, $\pm$25\textdegree{}, $\pm$40\textdegree{}, $\pm$60\textdegree{}; Set 3: $\pm$7.5\textdegree{}, $\pm$15\textdegree{}, $\pm$30\textdegree{}, $\pm$45\textdegree{}; Set 4: $\pm$2\textdegree{}, $\pm$4\textdegree{}, $\pm$17\textdegree{}, $\pm$27\textdegree{}) where $\pm$ indicates that the clamped feedback could be rotated clockwise (-) or counterclockwise (+). Given that eight perturbations within a set have a mean of zero, learning should not accumulate. The same demonstration trials (see Experiment 2) were included before the rotated clamped feedback block. 

Design of the in-person study (Tsay, Haith, et al., 2021): Participants (n = 12) performed reaching movements to the 90\textdegree{} target (straight ahead). There were 100 baseline reaching trials with veridical feedback. Each participant then completed four mini-blocks of perturbation trials (804 perturbation trials = 4 mini-blocks x 201 trials/mini-block). Mini-blocks were composed of either clamped, non-contingent rotation trials (0\textdegree{}, $\pm$4\textdegree{}, $\pm$8\textdegree{}) or target jump trials (0\textdegree{}, $\pm$4\textdegree{}, $\pm$8\textdegree{}; non-zero target jumps were not included in the analyses since this manipulation was outside the scope of the current study). Within each mini-block, there were 20 trials per condition presented in a random order (with the exception of 21 trials for 0\textdegree{} clamp x 0\textdegree{} target jump), and therefore 80 trials per clamp size x target jump combination across the entire experiment (and 84 trials for 0\textdegree{} clamp x 0\textdegree{} target jump). 

\subsection{Attention and instruction checks} It is difficult to verify if online participants fully attend to the task. To address this issue, we sporadically instructed participants to make specific key presses: “Press the letter “b” to proceed.” If participants failed the make an accurate key press, the experiment was terminated. These attention checks were randomly introduced within the first 50 trials of the experiment. We also wanted to verify that the participants understood the task, and in particular, understood in Experiments 2 and 3 that the angular position of the feedback was independent of the direction of their hand movement. To this end, we included one instruction check after the three demonstration trials: “Identify the correct statement. Press 'a': I will aim away from the target and ignore the white dot. Press 'b': I will aim directly towards the target location and ignore the white dot.” The experiment was terminated if participants failed to make the correct response (i.e., press “b”). The experiment was also terminated if the participant’s exhibited volatile behavior (i.e., exhibiting hand angle SD greater than 10\textdegree{} during baseline phases).

\subsection{Data analysis} The primary dependent variable of reach performance was hand angle, defined as the angle of the hand relative to the target when movement amplitude reached 6 cm from the start position (i.e., angle between a line connecting the start position to the target and a line connecting the start position to the hand). To aid visualization, the hand angle values for the groups (or trials in Experiment 3) with counterclockwise rotations were flipped, such that a positive hand angle corresponds to an angle in the opposite direction of the rotated feedback, the direction expected to result from learning.

Outlier responses were defined as trials in which the hand angle deviated by more than three standard deviations from a moving 5-trial window or if the hand angle was greater than 90\textdegree{} from the target. These outlier trials were excluded from further analysis since behavior on these trials could reflect attentional lapses or anticipatory movements to another target location (average percent of trials removed per participant: Experiment 1: 2.0 $\pm$ 1.0\%; Experiment 2: 1.5 $\pm$ 1.1\%; Experiment 3: 1.0 $\pm$ 0.7\%). Participants were excluded from analyses if more than 10\% of data points were removed. 11 participants were excluded using this criterion and were replaced with new participants. 

Experiments 1 and 2: The analysis protocol mimicked the two studies \cite{Bond2015-iu, Morehead2017-bf}. The mean heading angle for each movement cycle was calculated and baseline subtracted to evaluate adaptation relative to idiosyncratic movement biases. Baseline was defined as the last 5 cycles of the baseline block with veridical feedback (cycles 24 – 28). We evaluated three hand angle measures: Early adaptation, late adaptation, and aftereffect. Early adaptation was operationalized as the average mean hand angle over cycles 31 – 35 (cycles 3-7 of the rotation block). Late adaptation was defined as the mean hand angle over cycles 64 – 68 (cycles 35 – 40 of the rotation block, mimicking \cite{Bond2015-iu, Morehead2017-bf}). The aftereffect was operationalized as the average mean angle over the first 5 cycles of the no-feedback aftereffect block (cycles 79 – 83). 

All dependent measures were evaluated using an ANCOVA permutation test (R statistical packages: aovperm in the permuco package; 5000 permutations). This is a more robust test when the data are either normally or non-normally distributed \cite{Lehmann2010-ww}. Post-hoc pairwise permutation t-tests were performed (R statistical package: perm.t.test), and p values were Bonferroni correct to assess group differences.

Experiment 3: As our measure of trial-by-trial adaptation, we calculated the change in hand angle on trial n + 1 as a function of the rotation size on trial n for each trial. Means were then calculated for each clamp size, averaging over the clockwise and counterclockwise perturbations for a given size. These mean data were submitted to a linear regression to estimate the slope from the learning function for each individual (R statistical package: lm), with Rotation Size as the main effect. To ask whether these learning functions were sublinear, we compared each individual’s slope computed with all four rotation sizes, against the slope computed with the two smallest rotation sizes in their set. If adaptation was sublinear, then the slopes computing using all rotation sizes would be smaller in absolute magnitudes (less negative) than the slope computed using only small rotation sizes. 

\subsection{Implementing OnPoint on your own computer} The OnPoint package containing all the experiment code can be accessed and downloaded from:

(1) GitHub: https://github.com/alan-s-lee/OnPoint  \par
(2) Gorilla: https://gorilla.sc/openmaterials/111001 

The experiment is a webpage built using JavaScript, HTML, and CSS. The data collected from the experiment is stored via the Google Firebase database. The package also includes a .csv file that specifies various parameters of each trial (e.g., target location, target distance, rotation size) for a demo experiment (that researchers can try out at https://multiclamp-c2.web.app/) as well as a script that helps download data from the Firebase database. These files can be easily modified for researchers’ purposes. Other software requirements include installation of the Python3, NPM and Firebase packages (see GitHub). To recruit participants, researchers can use crowdsourcing platforms that allow the experiment to be shared via a hyperlink (i.e., website) to the experiment (e.g., AMT or Prolific).

\section*{acknowledgements}
All authors contributed to the study design. Testing, data collection, and data analysis were performed by J.S.T. All authors drafted the manuscript and approved the final version of the manuscript for submission.

\section*{conflict of interest}
None to disclose.

\printendnotes

\bibliography{sample}

\section{Supplemental Figure}

\begin{figure}[!htb]
\centering
\includegraphics[width=9cm]{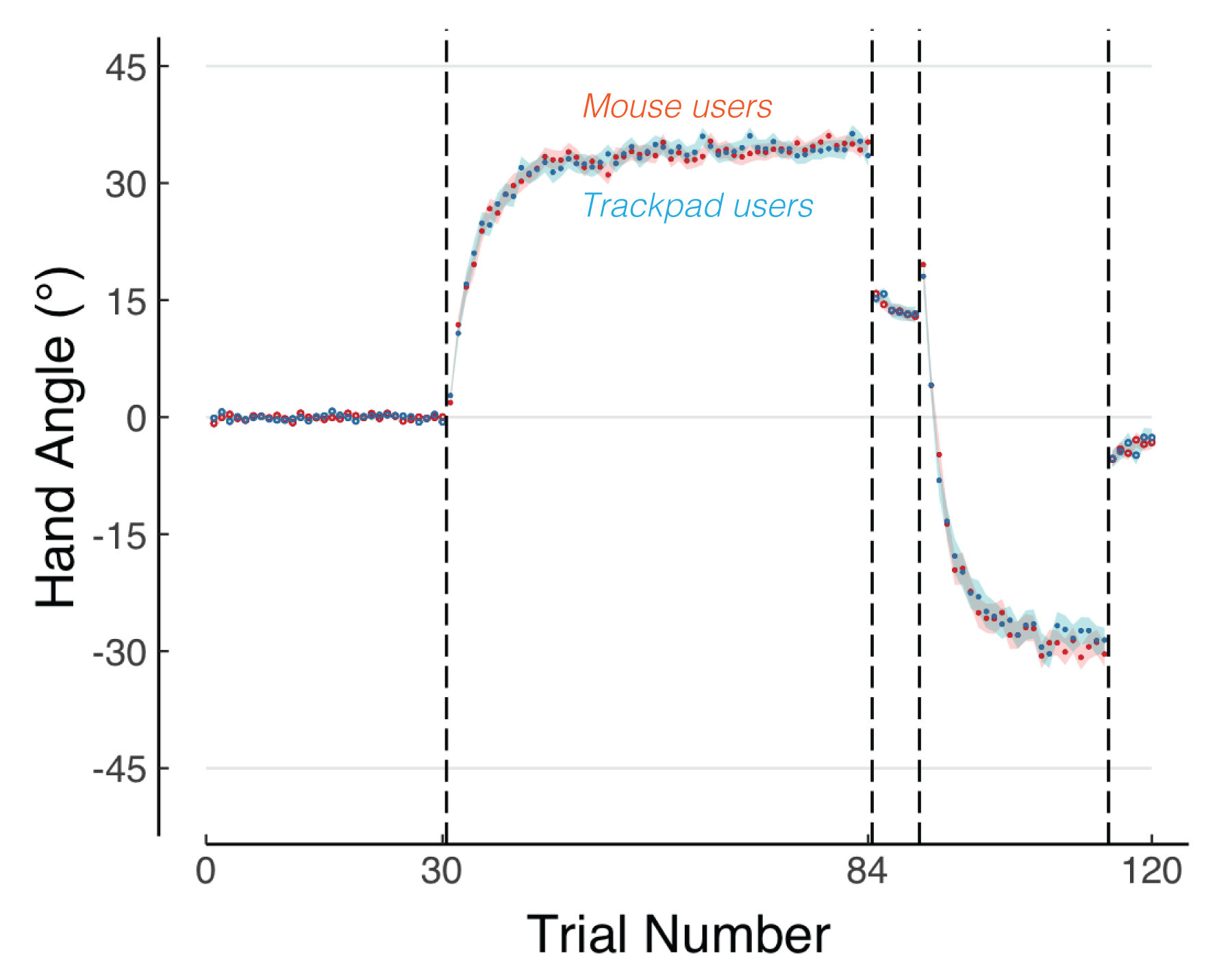}
\caption{\textbf{Motor adaptation is similar between trackpad users (n = 205) and mouse users (n = 225).} Participants completed five experimental blocks: a veridical feedback block (trials 1 - 30), a -45 deg rotation block (trials 31 - 84), a no feedback aftereffect block (trials 85 - 90), a 45 deg rotation block (trials 91 - 115), and a no feedback aftereffect block (trials 115 - 120).}
\end{figure}

\section{Supplemental Tables}

As an exploratory, post-hoc analysis, we evaluated key pairwise group comparisons between in-person and online settings. We performed pairwise t-tests with P values corrected for multiple comparisons (see captions). We have denoted significant group differences as shaded grey cells, operationally defined as P values being less than 0.05 and Bayes Factor ($BF_{01}$: in favor of the null) being less than 1 (Tables 1, 2, 3). 

While the results were qualitatively similar between in-person and online (see main text using ANCOVA regression analysis), this exploratory analysis revealed some quantitative differences. 

Experiment 1: Early adaptation was lower in the online setting compared to the in-person setting for 30\textdegree{}, 60\textdegree{}, and 90\textdegree{} groups, with the Bayes Factor providing weak to moderate support for the alternative hypothesis (Table 1, see how to interpret Bayes factor below). This result is consistent with the significant group x setting interaction observed in our regression analysis. Late adaptation was also lower for the 30\textdegree{} and 60\textdegree{} groups in the online setting, with the Bayes Factor providing moderate to strong support. Lastly, aftereffects were greater in the 30\textdegree{} and 60\textdegree{} group, with the Bayes factor providing strong support. 

Experiment 2: Data between experimental settings were similar for all phases (Table 2), with no significant differences detected. The Bayes Factor provides weak support for the null hypothesis.

Experiment 3: We compared the change in hand angle between settings for two clamp sizes: 4\textdegree{} online vs 4\textdegree{} in person, and 15\textdegree{} online (closest comparable rotation size) vs 16\textdegree{} in person (Table 3). We found that adaptation was only lower in the online group in the 4\textdegree{} clamp condition, with the Bayes Factor providing weak support for the alternative hypothesis. Nonetheless, the shape of the functions in Figs 3b-c were qualitatively similar, exhibiting a sublinear function bounded between $\pm$ 2.5\textdegree{}.

Interpreting Bayes Factor: The Bayes factor provides the ratio between the likelihood of two hypotheses (i.e., the likelihood that the data support the null hypothesis over the likelihood that the data support the alternative hypothesis). As such, a Bayes factor greater than 1 is an indication that the data favors the null hypothesis, with the magnitude reflecting the effect size. As a rule of thumb, Bayes factor values between 1 – 3, 3 – 10, and above 10 are considered to provide weak, moderate, and strong support for the null hypothesis, respectively \cite{Kass1995-dq, Lavine1999-fj}. Bayes factor values between 0.33 – 1, 0.1 – 0.33, and less than 0.1 are considered to provide weak, moderate, and strong support for the alternative hypothesis. 

\begin{table}[!htb]
\caption{\textbf{Experiment 1 hand angle analysis.} P values were corrected within each phase (i.e., 4 pairwise comparisons). Grey shaded regions denote groups where in-person differs to online, operationally defined as P values less than 0.05 and $BF_{01}$ less than 1.}

\begin{tabular}{|c|c|c|c|c|c|}
\headrow
Phase & \begin{tabular}{@{}c@{}}Rotation\\ Group \end{tabular} & \begin{tabular}{@{}c@{}}In-Person\\ Mean (SD) \end{tabular} & \begin{tabular}{@{}c@{}}Online\\ Mean (SD) \end{tabular} & P value & $BF_{01}$\\
\multirow{ 4}{*}{Early} & 15\textdegree{} & 9.5 (4.5) & 5.7 (4.2) & 0.12 & 0.37 \\
& \cellcolor{black!8}30\textdegree{} & \cellcolor{black!8} 24.1 (10.7) & \cellcolor{black!8} 12.1 (7.4) & \cellcolor{black!8}0.02 & \cellcolor{black!8}0.02\\
& \cellcolor{black!8}60\textdegree{} & \cellcolor{black!8}43.4 (22.5) & \cellcolor{black!8} 14.5 (32.3) & \cellcolor{black!8}0.01 & \cellcolor{black!8}0.26\\
& \cellcolor{black!8}90\textdegree{} & \cellcolor{black!8}79.9 (19.3) & \cellcolor{black!8}30.3 (38.9) & \cellcolor{black!8}<0.001 & \cellcolor{black!8}0.02\\
\hline  
\multirow{ 4}{*}{Late} & 15\textdegree{} & 12.7 (3.0) & 12.2 (4.3) & 1 & 2.74 \\
& \cellcolor{black!8}30\textdegree{} & \cellcolor{black!8}30.1 (5.7) & \cellcolor{black!8}21.3 (5.2) & \cellcolor{black!8}<0.001 & \cellcolor{black!8}0.00\\
& \cellcolor{black!8}60\textdegree{} & \cellcolor{black!8}60.4 (3.2) & \cellcolor{black!8}45.5 (15.5) & \cellcolor{black!8}<0.001 & \cellcolor{black!8}0.12\\
& 90\textdegree{} & 93.3 (4.9) & 89.5 (16.9) & 1 & 2.36\\
\hline  
\multirow{ 4}{*}{After} & 15\textdegree{} & 9.6 (4.0) & 8.1 (3.0) & 1 & 1.58 \\
& \cellcolor{black!8}30\textdegree{} & \cellcolor{black!8}9.9 (4.4) & \cellcolor{black!8}18.0 (7.2) & \cellcolor{black!8}<0.001 & \cellcolor{black!8}0.06\\
& \cellcolor{black!8}60\textdegree{} & \cellcolor{black!8}6.1 (3.9) & \cellcolor{black!8}17.5 (9.6) & \cellcolor{black!8}<0.001 & \cellcolor{black!8}0.03\\
& 90\textdegree{} & 6.9 (5.6) & 7.4 (7.0) & 1 & 2.80\\
\hline  

% Please only put a hline at the end of the table
\end{tabular}
\end{table}

\begin{table}[!htb]
\caption{\textbf{Experiment 2 hand angle analysis.} P values were corrected within each phase (i.e., 3 pairwise comparisons). Grey shaded regions denote groups where in-person differs to online, operationally defined as P values less than 0.05 and $BF_{01}$ less than 1.}

\begin{tabular}{|c|c|c|c|c|c|}
\headrow

Phase & \begin{tabular}{@{}c@{}}Rotation\\ Group \end{tabular} & \begin{tabular}{@{}c@{}}In-Person\\ Mean (SD) \end{tabular} & \begin{tabular}{@{}c@{}}Online\\ Mean (SD) \end{tabular} & P value & $BF_{01}$\\
\multirow{ 3}{*}{Early} & 7.5\textdegree{} & 5.2 (3.1) & 4.4 (3.0) & 1 & 2.42 \\
& 15\textdegree{} & 6.3 (3.2) & 3.4 (3.4) & 0.11 & 0.48\\
& 30\textdegree{} & 5.0 (3.4) & 5.3 (3.4) & 1 & 2.70\\
\hline  
\multirow{ 3}{*}{Late} & 7.5\textdegree{} & 13.4 (7.3) & 14.5 (4.3) & 1 & 2.55 \\
& 15\textdegree{} & 14.9 (10.1) & 14.2 (7.4) & 1 & 2.74\\
& 30\textdegree{} & 12.1 (4.9) & 15.3 (3.8) & 0.33 & 0.61\\
\hline  
\multirow{ 3}{*}{After} & 7.5\textdegree{} & 12.1 (8.1) & 14.5 (4.2) & 1 & 1.75 \\
& 15\textdegree{} & 11.8 (9.1) & 11.9 (6.1) & 1 & 2.78\\
& 30\textdegree{} & 10.5 (4.9) & 15.3 (3.8) & 0.06 & 0.15\\
\hline  
\end{tabular}
\end{table}

\begin{table}[!htb]
\caption{\textbf{Experiment 3 hand angle analysis.} P values were corrected within each phase (i.e., 2 pairwise comparisons). Grey shaded regions denote groups where in-person differs to online, operationally defined as P values less than 0.05 and $BF_{01}$ less than 1.}

\begin{tabular}{|c|c|c|c|c|c|}
\headrow

\begin{tabular}{@{}c@{}}In-Person\\ Clamp Size \end{tabular}&\begin{tabular}{@{}c@{}}Online\\ Clamp Size \end{tabular} & \begin{tabular}{@{}c@{}}In-Person\\ Mean (SD) \end{tabular} & \begin{tabular}{@{}c@{}}Online\\ Mean (SD) \end{tabular} & P value & $BF_{01}$\\
4\textdegree{} & 4\textdegree{} & 1.1 (0.3) & 0.7 (0.8) & 0.02 & 0.87 \\
\hline
16\textdegree{} & 15\textdegree{} & 1.8 (0.3) & 1.7 (1.1) & 1 & 2.66\\
\hline  
\end{tabular}
\end{table}

\end{document}